\documentclass[11pt]{amsart}
\usepackage[english]{babel}
\usepackage[utf8]{inputenc}
\usepackage{csquotes}

\usepackage{geometry}
\usepackage{amssymb,amsmath,amscd,amsfonts,amsthm,bbm,xcolor}

\newtheorem{theorem}{Theorem}
\newtheorem{conjecture}[theorem]{Conjecture}

\usepackage[colorlinks=true]{hyperref}

\usepackage{amsaddr}
\usepackage{braket}
\usepackage{graphicx}
\usepackage{subcaption}
\usepackage{tabularx}
\usepackage{xcolor}

\usepackage[
  backend=biber,
  style=numeric,
  sorting=none,
  giveninits=true,
  doi=false,
  url=false,
  isbn=false,
  eprint=false,
  autolang=other,
  maxnames=99,
  maxcitenames=99
]{biblatex}

\renewbibmacro{in:}{}
\AtEveryBibitem{%
  \clearfield{language}%
}

\DeclareFieldFormat[article,inbook,incollection,inproceedings]{title}{#1}

\usepackage{setspace}

\usepackage[mathlines]{lineno}

\begin{document}

\nolinenumbers 

\title{Approximate message passing for block-structured ecological systems}
\author{Maxime Clenet~$^{1,2}$}
\address{Département de biologie, Université de Sherbrooke, Canada}

\author{Mohammed-Younes GUEDDARI~$^{1,3}$}
\address{Laboratoire d’informatique Gaspard Monge, UMR 8049, Université Gustave Eiffel, France}

%
%

%
\begin{abstract}
Ecological interaction networks are rarely homogeneous: species naturally form communities with distinct interaction structures, resulting in block-structured variance and correlation profiles in the interaction matrix. We study the equilibrium properties of generalized Lotka–Volterra systems whose interaction matrices are random and non-symmetric with variance and correlation profiles. Based on recent advances in approximate message passing (AMP) for heterogeneous and correlated random matrices, we derive a set of self-consistent fixed-point equations that, in the large-$n$ limit, characterize the equilibrium abundance distribution. In particular, we show that this limiting distribution is an explicit mixture of truncated Gaussians, driven by the variance and correlation profiles. We then illustrate the ecological implications of this result through three applications involving two interacting communities. First, we show that local changes in the correlation profile within a single community induce system-wide responses in species persistence, revealing the non-local nature of persistence dynamics. Second, we find that communities dominated by mutualistic or competitive interactions are more robust to increasing inter-community coupling, whereas communities structured by predator–prey interactions are more prone to collapse. Third, we demonstrate that asymmetric interaction variance alone, in the complete absence of correlation, can generate feedback loop between communities.

\end{abstract}

\maketitle
\begin{raggedleft}
\end{raggedleft}
\setcounter{footnote}{1}
\footnotetext{~M.C. and M-Y.G. contributed equally to this work.}
\setcounter{footnote}{2}
\footnotetext{~Present address: CEFE, Université de Montpellier, CNRS, Montpellier, France}
\setcounter{footnote}{3}
\footnotetext{~To whom correspondence may be addressed. Email: mohammed-younes.gueddari@univ-eiffel.fr}

%

\newpage

\section{Introduction}

\subsection{Motivations}

Ecological communities are characterized by biotic interactions, such as competition, mutualism, and predation \cite{vellend_theory_2016}. These interactions are often represented as a network of species interactions \cite{montoya_ecological_2006}. Together, these interactions influence community composition and long-term persistence \cite{mougi_diversity_2012,landi_complexity_2018}. For example, interaction matrix often exhibit modular or block structures that reflect functional groups, trophic levels, or spatial organization. This type of pattern tends to promote species persistence \cite{guimera_origin_2010,stouffer_compartmentalization_2011}.

A fundamental question is how the structure of species interactions influences the stability of ecological systems. Stability is defined as the ability of a system to return to equilibrium after a disturbance. The analysis of interactions in the vicinity of equilibrium as a high-dimensional dynamical system resulted in the concept of the community matrix \cite{may_will_1972}. The community matrix is defined as the Jacobian of the system evaluated at equilibrium. Each entry of the community matrix quantifies the direct effect of small changes in the abundance of one species on the growth rate of another. The community matrix was modeled as a random matrix \cite{may_will_1972}. This approach has provided a means to study how the statistical properties of interactions, such as mean strength, variance, and correlation, influence the stability of ecological systems at equilibrium \cite{may_will_1972,allesina_stability_2012}. In particular, May \cite{may_will_1972} suggested that the subdivision of ecosystems into multiple, weakly connected communities could have a stabilizing effect, a prediction later supported by both theoretical and empirical analyses \cite{thebault_stability_2010,stouffer_compartmentalization_2011,allesina_predicting_2015}.

One major limitation of the community matrix paradigm is its assumption that the system is at equilibrium. The next step is to understand how the structure of the interaction matrix affects species abundance dynamics. This challenge has motivated the development of analytical models that explicitly link the statistical properties of interaction matrices to the dynamics of large systems of interacting species. In particular, the combination of Lotka–Volterra (LV) dynamics \cite{lotka_elements_1925,volterra_fluctuations_1926} with random matrix theory has become a cornerstone of theoretical ecology \cite{goh_feasibility_1977,akjouj_complex_2024}. The objective was similar to that of the community matrix paradigm: to understand how the statistical properties of the interaction matrix influence stability and persistence within the LV model \cite{bunin_ecological_2017,servan_coexistence_2018}.
\newline

Among these statistical properties, correlations between pairwise interactions play a central role \cite{barbier_generic_2018,poley_generalized_2023}. A negative correlation between two species corresponds to a predator-prey interaction, whereas a positive correlation tends to correspond to a mutualistic or competitive interaction. Clenet \textit{et al.} \cite{clenet_equilibrium_2022} showed that correlation does not affect the feasibility threshold, i.e. the boundary at which all species maintain strictly positive equilibrium abundances. Beyond the feasible threshold where some species can go extinct, correlation has a significant impact on species persistence \cite{bunin_ecological_2017}. Negative correlations tend to enhance persistence, while positive correlations reduce it relative to the uncorrelated case. These findings are consistent with classical results of the community matrix, where predator–prey interactions are known to have a stabilizing effect on community dynamics \cite{allesina_stability_2012,tang_correlation_2014,thebault_stability_2010}.

While correlations between pairwise interactions have been a central statistical feature in previous studies, the impact of structured correlation profiles on equilibrium properties has received far less attention. Allowing for a block-structured profile of variance and correlation introduces an additional level of interaction structure, in which species naturally form communities \cite{baskerville_spatial_2011,grilli_modularity_2016}. Clenet \textit{et al.} \cite{clenet_impact_2024} investigated the effect of block variance profiles, without any correlations, on stability, feasibility and the attrition phenomena where some species go extinct. 

Recently, a powerful theoretical tool has emerged in the study of high-dimensional systems, including theoretical ecology: \textit{Approximate Message Passing} (AMP). Originally introduced in the context of compressed sensing and sparse signal recovery \cite{Donoho_2009, Donoho_2009_1, Donoho_2009_2, bayati2011dynamics}, AMP provides an iterative scheme with powerful statistical properties, often used to analyze large random systems and high-dimensional inference problems. For a given random matrix $A$, the AMP iteration typically takes the form
\[
x^{t+1} = A\, h_t(x^t) - \{\text{Onsager correction term}\} \, ,
\]
where the specific form of the corrective term depends on structural properties of $A$, such as its variance or correlation profile. AMP has since found numerous applications, notably in the study of spiked random matrix models and rank-one signal estimation problems \cite{Rangan2012IterativeEO, javanmard2013state, deshpande2014}.

In the context of theoretical ecology, AMP was first introduced by \cite{akjouj_equilibria_2024} to analyze the high-dimensional equilibria of the LV system when the interaction matrix follows a Gaussian Orthogonal Ensemble (GOE). This approach was later extended in \cite{gueddari_elliptic_2025} to the case of elliptic random matrices, where symmetric entries are correlated with a constant correlation coefficient. More recently, motivated by the modeling of structured ecological communities, new AMP algorithms have been developed \cite{hachem_approximate_2024, gueddari_approximate_2026}, adapted to random matrices with heterogeneous variance and correlation profiles. These developments open the way for the rigorous study of structured interaction matrices, where species form communities with distinct statistical interaction patterns.

\subsection{Model and assumptions}
In this work we study the application of sparse and correlated AMP algorithms \cite{gueddari_approximate_2026} to the analysis of equilibrium of LV systems. 

Let $r\in \mathbb{R}_{*+}^n$ be the intrinsic growth rates vector of the species and consider an $n\times n$ dimensional non-symmetric random matrix $\Sigma \in \mathbb{R}^{n\times n}$ which represents the interaction matrix. The dynamics of species abundances are governed by the LV system
\begin{equation*}
    \frac{d}{dt} u(t) = u(t) \odot \left(r - \left(I_n - \Sigma\right) u(t)\right), \quad t\geq 0 \, .
\end{equation*}
Here $u(t)\in \mathbb{R}^n$ is the vector of species abundances at time $t$ and $\odot$ stands for the component-wise Hadamard product. The equilibrium $u_\star\in \mathbb{R}^n$ (as $t\to \infty$) satisfies the following system of equations
\begin{equation*}
    \label{eq:equilibrium}
    u_\star \odot \left(r - \left(I_n - \Sigma\right) u_\star\right) = 0 \, .
\end{equation*}
In particular, each component of the vector $u_\star$ is either zero (extinct species) or positive.
And in addition, the equilibrium is required to satisfy the Lyapunov stability condition, which in theoretical ecology is equivalent to the non-invadability condition,
\begin{equation}
\label{eq:noninvadability}
r-\left(I_n-\Sigma\right)u_\star \leq 0,
\end{equation}
where the inequality is understood component-wise. This condition ensures that species absent from the equilibrium cannot increase from arbitrarily small introductions.

Our objective is to describe the asymptotic behavior (as $n\to\infty$) of the empirical measure $\mu_n^{u^\star}$ under a specific setup
$$\mu_n^{u^\star} := \frac{1}{n}\sum_i \delta_{u^\star_i}\, . $$

\subsection{Outline of the article} Section \ref{sec:main_result} introduces the block-structured interaction model and specifies the correlation profile and statistical assumptions used throughout the article. Theorem \ref{thm:main_block} establishes the existence of a limiting equilibrium and characterizes the abundance distribution as the solution of a set of self-consistent fixed-point equations. We also state Conjecture \ref{conj:general}, which extends the result to a general interaction model with arbitrary correlation and variance profiles. The proof of Theorem \ref{thm:main_block}, including the derivation of the AMP equations and the technical arguments establishing their asymptotic validity in the block-structured setting, can be found in Appendix \ref{app:proof}.
Section \ref{sec:applications} presents three applications illustrating how variance and block-structured correlation profile shape species persistence and equilibrium abundance distributions. First, we show that modifying the correlation profile while keeping the variance fixed affects persistence at the global rather than local level: introducing structure within a single community impacts the persistence of all species. Second, we find that communities dominated by mutualistic or competitive interactions are more robust to increases in inter-community variance, whereas communities structured by predator–prey interactions are more sensitive and prone to collapse. Third, we uncover an emergent feedback loop between communities, whereby a decline in persistence in one community can trigger a cascading collapse in another.

\section{Main result}
\label{sec:main_result}

\subsection{Model setup}

Let $T = (\tau_{ij}) \in [-1,1]^{n \times n}$ be a matrix of correlation coefficients. 
We consider a $T$-correlated random matrix $X \in \mathbb{R}^{n \times n}$, meaning that
\[
\operatorname{Corr}(X_{ij}, X_{ji}) = \tau_{ij} \, .
\]
Let $V = (v_{ij}) \in \mathbb{R}^{n \times n}$ be a deterministic matrix satisfying suitable assumptions. 
The interaction matrix $\Sigma$ is defined by
\begin{equation}
\label{eq:matrixModel}
    \Sigma = V^{\odot 1/2} \odot X \, ,
\end{equation}
where $\odot$ denotes the Hadamard (entry-wise) product.

We focus on the case where $\Sigma$ exhibits a $K \times K$ block structure, with $K$ fixed (independent of $n$). 
Let $S = (s_{ij}^2)_{1 \le i,j \le K}$ be a matrix with strictly positive entries (not necessarily symmetric), 
and let $R = (\rho_{ij})_{1 \le i,j \le K}$ be a symmetric matrix with entries in $[-1,1]$.

We define variance and correlation profiles $V,T \in \mathbb{R}^{n \times n}$ structured into $K \times K$ blocks as follows:
the $(i,j)$-th block of $V$ (respectively $T$) has constant value $s_{ij}^2 / n$ (respectively $\rho_{ij}$). 
The block sizes are $\alpha_1 n, \dots, \alpha_K n > 0$, where
\[
\alpha_1 + \cdots + \alpha_K = 1 \, .
\]

\subsection{Main result}

For $u,v \in \mathbb{R}^n$, we use the notation $uv := u \odot v = (u_i v_i)_{i=1}^n.$ For a matrix $A$, we denote by $\|A\|_\infty$ its maximal row sum. We are now ready to state the main theorem.

\begin{theorem}
\label{thm:main_block}
Assume that $\|S\|_\infty < \frac{1}{8}$ and consider the interaction matrix $\Sigma$ defined above with block-structured variance and correlation profiles.\\
Assume further that \(r \in \mathbb{R}^n\) is a random vector such that, for each $k \in [K]$, the restriction of $r$ to the $k$-th block (of dimension $\alpha_k n$) has an empirical measure that converges, in Wasserstein distance (denoted as $\boldsymbol{d}_2$), to the law of a random variable $\bar{r}_k$.
\\
Then the empirical measure of the equilibrium vector $u^\star$ satisfies
\[
\boldsymbol{d}_2\big( \mu_n^{u^\star}, \mu_\infty \big)
\;\xrightarrow[n \to \infty]{\mathbb{P}}\; 0,
\]
where
\begin{equation}
\label{eq:truncated_gaussian}
\mu_\infty = \sum_{k=1}^K \alpha_k \mu_k,
\qquad
\mu_k = \mathcal{L}\!\left(
\frac{1}{\delta_k^\star} \big( \sigma_k^\star \xi_k + \bar{r}_k \big)_+
\right),
\end{equation}
and $\xi_1,\dots,\xi_K \stackrel{i.i.d.}{\sim} \mathcal{N}(0,1)$. The vector $\big(\delta_1^\star, \cdots, \delta_K^\star, \sigma_1^\star, \cdots, \sigma_K^\star \big)$ is a solution of the following system of $2K$ equations:
\begin{equation}
\label{FPK}
\left\{
\begin{aligned}
1 &= \delta_i 
+ \sum_{j=1}^K 
\alpha_j \rho_{ij} s_{ij} s_{ji}
\frac{\mathbb{P}\!\left( \sigma_j \xi + \bar{r}_j \ge 0 \right)}{\delta_j},
& \quad i=1,\dots,K, \\[0.3cm]
\sigma_i^2 &= 
\sum_{j=1}^K 
\alpha_j s_{ij}^2
\frac{\mathbb{E}\!\left( \sigma_j \xi + \bar{r}_j \right)_+^2}{\delta_j^2},
& \quad i=1,\dots,K,
\end{aligned}
\right.
\end{equation}
with $\xi\sim \mathcal{N}(0,1).$
\end{theorem}

\begin{figure}[t]
    \centering
    \includegraphics[width=0.85\textwidth]{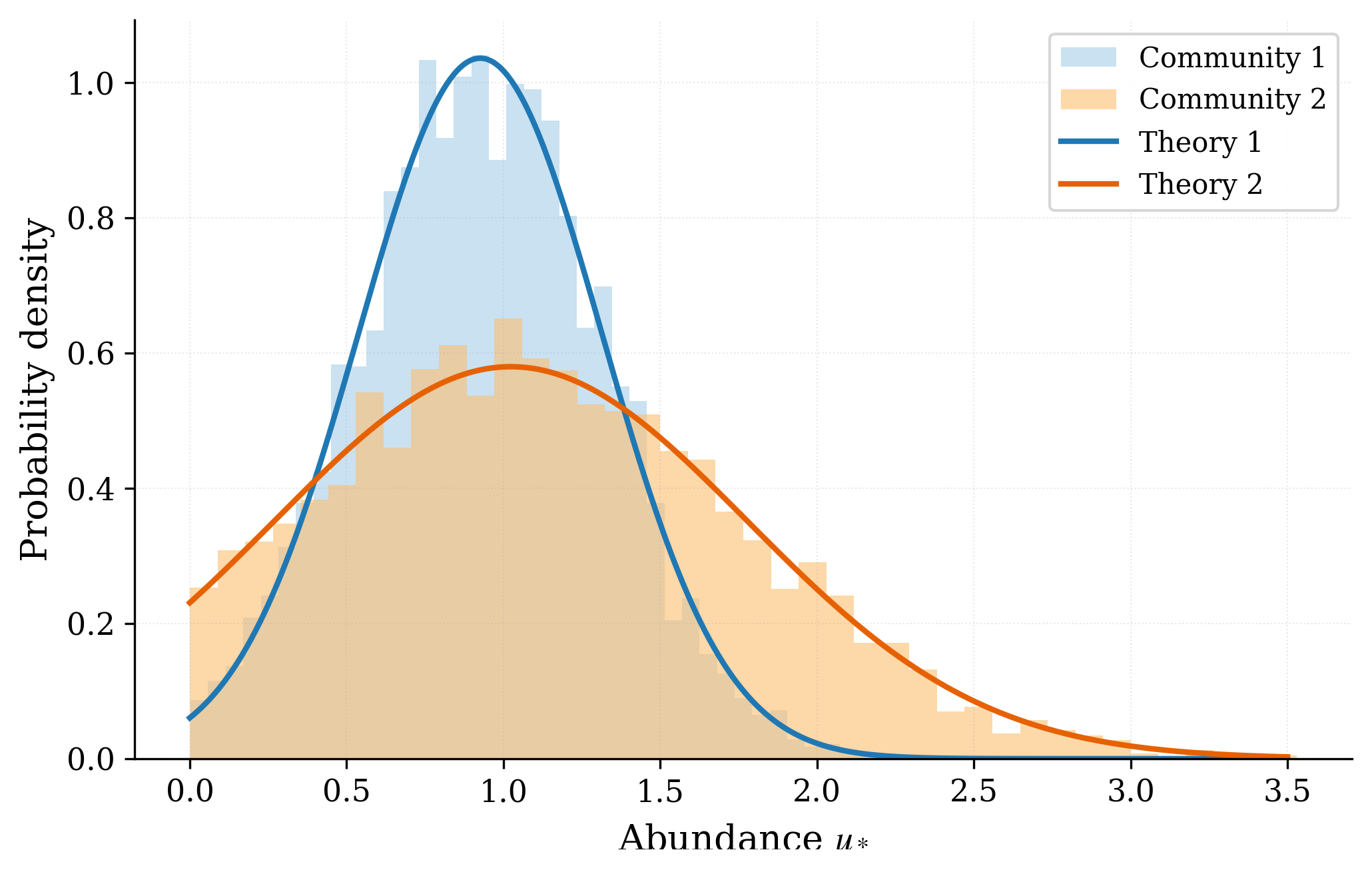}
    \caption{Empirical equilibrium abundance distributions (histograms) compared with the truncated normal densities predicted by the fixed-point equations \eqref{eq:truncated_gaussian} (dashed lines). Colors denote the two blocks of the interaction matrix,
$S = \begin{pmatrix}0.25 & 0.0625 \\ 0.0625 & 0.64\end{pmatrix},
\quad
R = \begin{pmatrix}-0.9 & 0.8 \\ 0.8 & 0\end{pmatrix},
\quad
\alpha = (0.5, 0.5),
\quad
n = 10^4$.
Blue and red correspond to communities 1 and 2, respectively.}
    \label{fig:lcp_mixture}
\end{figure}

Figure \ref{fig:lcp_mixture} compares the empirical distribution and the theoretical truncated Gaussian mixture \eqref{eq:truncated_gaussian}. The close match between the histograms and the theoretical density distributions confirms both the accuracy of the asymptotic prediction and the validity of the block-structured model, even beyond the condition $\|S\|_\infty < \frac{1}{8}$.

\subsection{Example} 
In the case of an elliptic matrix with a uniform variance profile, i.e. if $$V = s^2\,\frac{11^\top}{n}\in \mathbb{R}^{n\times n}\quad \text{ and }\quad T=(\tau_{ij}) = (\rho)\in [-1, 1]^{n\times n},$$ the system of equations \eqref{FPK} reduces to a system of only $2$ equations:
\begin{equation}
    \label{eq:FPE5}
    \left\{
    \begin{aligned}
        1 &= \delta +\rho s^2 \frac{\mathbb{P}\left[\sigma \xi + r \geq 0\right]}{\delta},\\
        \sigma^2 &= \frac{s^2}{\delta^2}\mathbb{E}\left(\sigma \xi + r\right)_+^2.
    \end{aligned}
    \right.
\end{equation}
This is exactly the system of equations that we can find in \cite{clenet_equilibrium_2022} (in the case of $\rho=0$ and a non-centered random matrix), \cite{akjouj_equilibria_2024} (in the case of $\rho=1$) and in \cite{gueddari_elliptic_2025} (general $\rho$).

The result, can then be formulated in the following way
$$ \boldsymbol{d}_2( \mu_n^{u^\star},\mu) \stackrel{\mathbb{P}}{\underset{n\to\infty}{\longrightarrow}} 0\,.  $$
where $\mu$ is given by 
$$\mu = \mathcal{L}\left(\frac{1}{\delta}\left(\sigma \xi+r\right)_+\right), \quad \xi \sim \mathcal{N}(0,1) \, .$$
The couple $(\sigma, \delta)$ is the solution of the system \eqref{eq:FPE5}.

\subsection{The general case of non structured matrices}

As will be explained in the outline of the proof (see Appendix~\ref{app:proof}), 
one can implement an AMP algorithm for the general interaction matrix model, 
without imposing any block structure. In fact, in \cite{gueddari_approximate_2026}, 
an AMP algorithm was developed for matrices with arbitrary variance and correlation profiles.

We therefore formulate the following conjecture regarding the limiting behavior 
of the empirical distribution of the equilibrium in this general setting. 
The main ingredient currently missing to elevate this conjecture to a theorem 
is a proof of the existence and uniqueness for the associated system of fixed-point equations.

This conjecture was proven in the case of a symmetric, possibly sparse variance profile without correlations in the work of Hachem \cite{hachem_approximate_2024}. The main difficulty in our setting comes from the presence of an arbitrary correlation structure.

\begin{conjecture}
\label{conj:general}
    Let $\mathcal{V} :=\left(V\odot V^\top\right)^{\odot 1/2}\odot T = (\sqrt{v_{ij}v_{ji}}\tau_{ij})$. Let $(p, \zeta)\in \mathbb{R}^n\times \mathbb{R}^n$ be the unique solution of the following system of $2n$ equations:
    \begin{equation}
    \label{eq:FPE2}
    \begin{cases}
        p &= V\operatorname{diag}(1+\zeta)^2\mathbb{E}\left(\sqrt{p} \xi + r\right)_+^2\\
        \zeta &= \operatorname{diag}(1+\zeta)\mathcal{V}\operatorname{diag}(1+\zeta) \mathbb{P}\left[\sqrt{p}\xi + r\geq 0\right]\, , 
    \end{cases} 
\end{equation}
where $\xi\sim\mathcal{N}_n(0,I_n)$. Let $Y\in \mathbb{R}^n$ be the Gaussian vector defined as 
$$ Y = \left(1+\zeta\right)\left(\sqrt{p}\xi + r\right), \quad \xi\sim\mathcal{N}_n(0,I_n).$$
We denote by $Y_i$ the $i$-th component of $Y$. Let $\theta \sim \mathcal{U}([n])$, finally consider the following \textit{deterministic} probability measure $\mu_n$ defined as 
$$ \mu_n = \mathcal{L}\left(\left(Y_\theta\right)_+\right) \, , $$
then we have the following result :
$$ \boldsymbol{d}_2( \mu_n^{u^\star},\mu_n) \stackrel{\mathbb{P}}{\underset{n\to\infty}{\longrightarrow}} 0. $$
\end{conjecture}

Note that the system of equations \eqref{eq:FPE2} reduces to the fixed-point system 
\eqref{FPK} when the interaction matrix is assumed to have a block structure.

The condition $\|S\|_\infty < \tfrac{1}{8}$ in Theorem \ref{thm:main_block} should be read as a sufficient condition ensuring the existence argument, not as a sharp threshold. As Figure~\ref{fig:lcp_mixture} illustrates, the predicted mixture remains accurate well beyond this bound. Accordingly, the applications of Section~\ref{sec:applications}, which are all set in the block-structured regime with $K=2$ and are thus governed by the system~\eqref{FPK}, fall within the framework of Theorem~\ref{thm:main_block} at the level of the equations, even though several of them use variance profiles with $\|S\|_\infty > \tfrac{1}{8}$ and therefore rely on this numerically validated extension rather than on the proven regime (see Appendix \ref{app:conj_num} for another example).

\section{Applications}
\label{sec:applications}

The main analytical results were presented using the AMP and summarized in Theorem~\ref{thm:main_block}. In this section, we will illustrate how these results can be interpreted within ecologically meaningful scenarios. To do so, we will focus on the case of two interacting communities ($K = 2$). This minimal yet rich setting allows us to explore how the structured variance and correlation profile of the interaction matrix influences the equilibrium distribution of species persistence and abundance.

In this two communities setting, the interaction matrix is defined by a variance profile
$S \in \mathbb{R}^{2 \times 2}$ and a correlation profile
$R \in \mathbb{R}^{2 \times 2}$, given by:
$$
S = \begin{pmatrix}
s_{11}^2 & s_{12}^2 \\
s_{21}^2 & s_{22}^2
\end{pmatrix}, 
\quad 
R = \begin{pmatrix}
\rho_{11} & \rho_{12} \\
\rho_{21} & \rho_{22}
\end{pmatrix}.
$$
Here, $s_{ij}^2$ denotes the variance (strength) of interaction coefficients from block $j$ to block $i$, and $ \rho_{ij} \in [-1,1]$ their correlation. We set $r=1$ and the size of the community is the same $\alpha_1 = \alpha_2 = \frac{1}{2}$.

By solving the fixed-point system~\eqref{FPK}, we obtain two sets of couples, $(\delta_1, \sigma^2_1)$ and $(\delta_2, \sigma^2_2)$, associated to communities 1 and 2 respectively. Here, $\hat{\sigma}^2_k = \frac{\sigma^2_k}{\delta^2_k}$ denotes the variance of species abundances at equilibrium in community $k$. We can recover the proportion of species that persist in each community by computing 
$$
\gamma_k = \mathbb{P}\left[\sigma_k \xi + 1 \geq 0\right] \quad\forall k\in \{1,2\}\, , \quad \xi\sim\mathcal{N}(0,1) \, .
$$  

Through three applications, we use this framework to examine the ecological implications of structured interaction matrices, each highlighting a distinct mechanism: (i) the effect of correlation profile on global persistence; (ii) the impact of increasing inter-community interaction when the communities have opposing interaction types; and (iii) feedback effects arising purely from variance asymmetry in the absence of correlation.

\subsection{Effect of correlation profile}

To assess the influence of local correlation on species abundance variance $\hat{\sigma}^2_k$ and persistence $\gamma_k$ at equilibrium, we considered the case of a homogeneous interaction variance profile across all blocks (i.e., $s^2_{ij} = 0.25$ for all $i,j$). Correlation was introduced locally by varying the intra-community correlation $\rho_{11} \in [-1,1]$ within community 1, while keeping community 2 and all inter-community interactions uncorrelated. 

\begin{figure}[ht]
    \centering
    \begin{subfigure}[b]{0.48\textwidth}
        \centering
        \includegraphics[width=\textwidth]{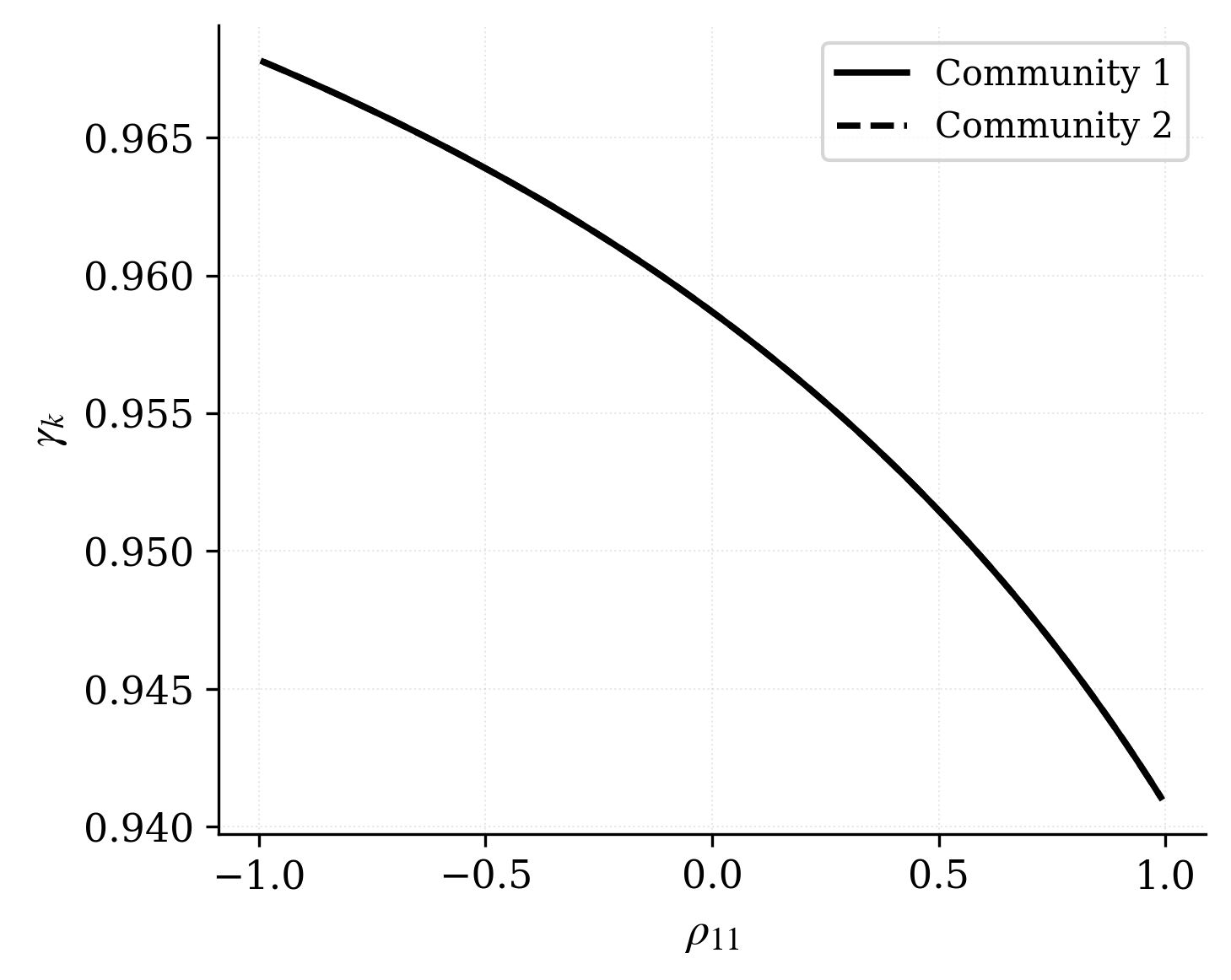}
        \caption{Proportion of persistent species in each community as a function of intra-correlation of community 1 $\rho_{11}$.}
        \label{fig:2a}
    \end{subfigure}
    \hfill
    \begin{subfigure}[b]{0.48\textwidth}
        \centering
        \includegraphics[width=\textwidth]{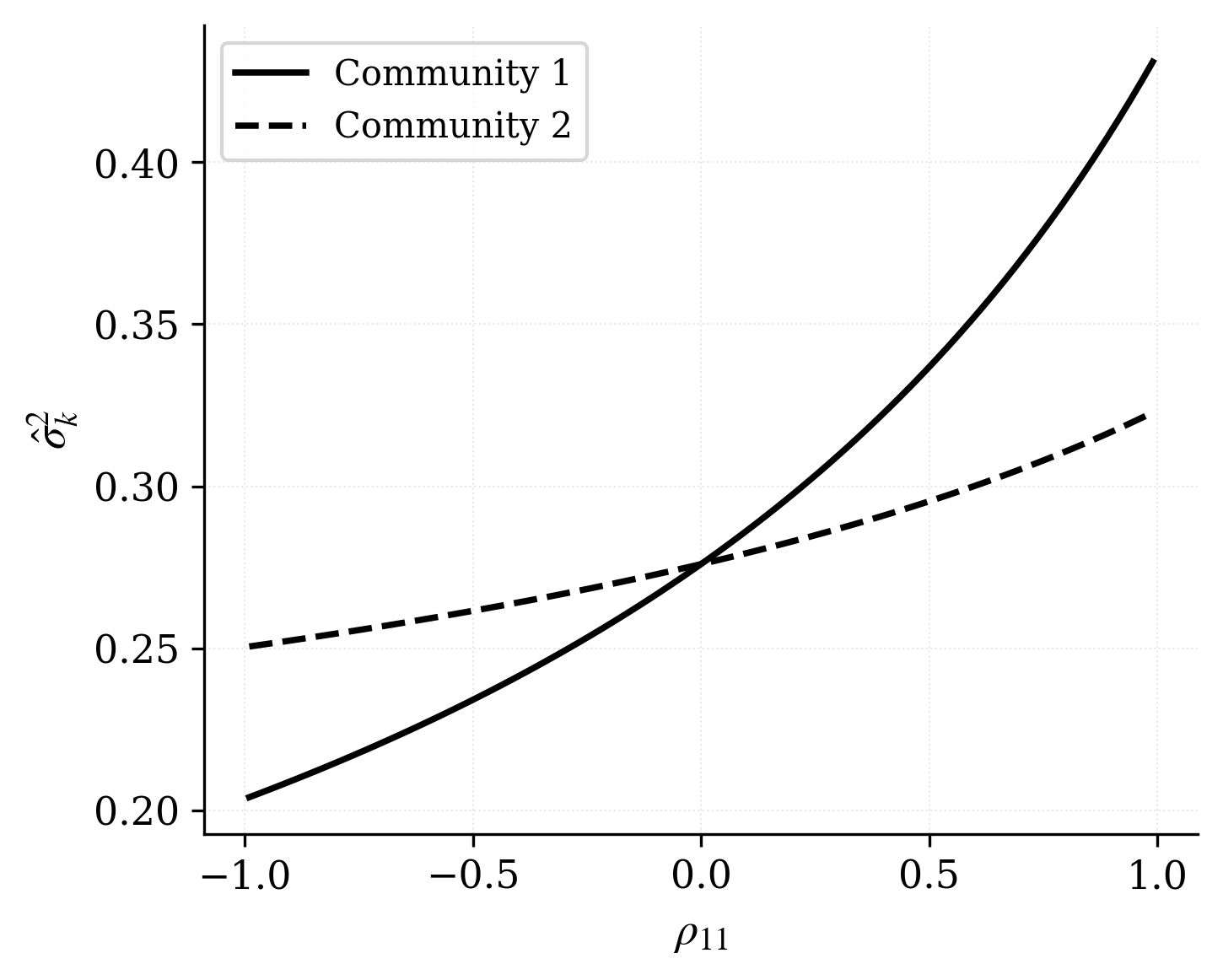}
        \caption{Equilibrium abundance variance in each community as a function of intra-correlation of community 1 $\rho_{11}$.}
        \label{fig:2b}
    \end{subfigure}
\caption{
Effects of correlation $\rho_{11}$ in community 1 on the properties of species abundance distribution.  
(a) Solid and dashed lines show the proportion of persisting species in each community as a function of the intra-community correlation $\rho_{11}$ in community 1.  
(b) Solid and dashed lines show the variance of equilibrium abundances in each community, under the same variation of $\rho_{11}$. While persistence responds globally and symmetrically, abundance variance exhibits a more localized and asymmetric response to the correlation $\rho_{11}$.
}
\label{fig:2}
\end{figure}

We found that increasing the correlation $\rho_{11}$ of community 1 affected the overall persistence of the system in line with known theoretical results \cite{clenet_equilibrium_2022}: negative correlation ($\rho_{11} < 0$) tends to increase persistence, whereas positive correlation ($\rho_{11} > 0$) reduced it. Interestingly, this change in persistence was evenly distributed across both communities, i.e. $\gamma_1 = \gamma_2, \forall \, \rho_{11} \in [-1,1]$. In other words, although correlation was introduced exclusively in community 1, the resulting species extinctions did not show a localized pattern, but instead reflected a global adjustment at the level of the two communities (Fig.~\ref{fig:2a}).

In contrast, the variance of species abundances at equilibrium responded asymmetrically (Fig.~\ref{fig:2b}). As $\rho_{11}$ increased, the variance rose sharply in community 1 and more moderately in community 2. This suggests a partial decoupling between the mechanisms governing species persistence, which respond globally and symmetrically and those shaping abundance variance, which respond more locally and asymmetrically. This emergent pattern demonstrates a type of collective compensation in which local structural changes redistribute ecological variability throughout the system.

These results demonstrate that a correlation profile localized within a single community reshapes persistence at the system-wide level rather than within the perturbed community alone. Negative correlation increases persistence in both communities, while positive correlation decreases it by an identical amount ($\gamma_1 = \gamma_2$). The block-specific signature of the perturbation does not appear in persistence but in the abundance variance $\hat{\sigma}^2_k = \sigma_k^2/\delta_k^2$, which responds asymmetrically: the correlation enters $\delta_1$ while leaving $\delta_2 = 1$ unchanged, so that $\hat{\sigma}^2_1$ is amplified relative to $\hat{\sigma}^2_2$ even though $\sigma_1^2 = \sigma_2^2$ (Fig.~\ref{fig:2b}). The equalization $\gamma_1 = \gamma_2$ is a direct consequence of the symmetric inter-community variance ($s^2_{12} = s^2_{21}$), which forces $\sigma_1^2 = \sigma_2^2$; it is therefore specific to this balanced setup and should not be read as a general decoupling between correlation and persistence. Application~2 confirms this caveat: once the variance profile becomes asymmetric, the sign of the intra-community correlation differentiates the robustness of the two communities.

\subsection{Opposing correlation structure}

In this second application, we explored how intra-community correlation structures modulate persistence under increasing inter-community interaction variance. Specifically, we considered two distinct communities of equal size, each with homogeneous interaction variances ($s^2_{11} = s^2_{22} = 0.25$), but opposing correlation patterns: a strong positive correlation in community 1 ($\rho_{11} = 0.8$) and a strong negative correlation in community 2 ($\rho_{22} = -0.8$). We progressively increased the variance of inter-community interactions ($s^2 = s^2_{12} = s^2_{21}$) and quantified the resulting effects on species persistence and the variance of equilibrium abundances ($\hat{\sigma}^2$) within each community.

\begin{figure}[ht]
    \centering
    \begin{subfigure}[b]{0.32\textwidth}
        \centering
        \includegraphics[width=\textwidth]{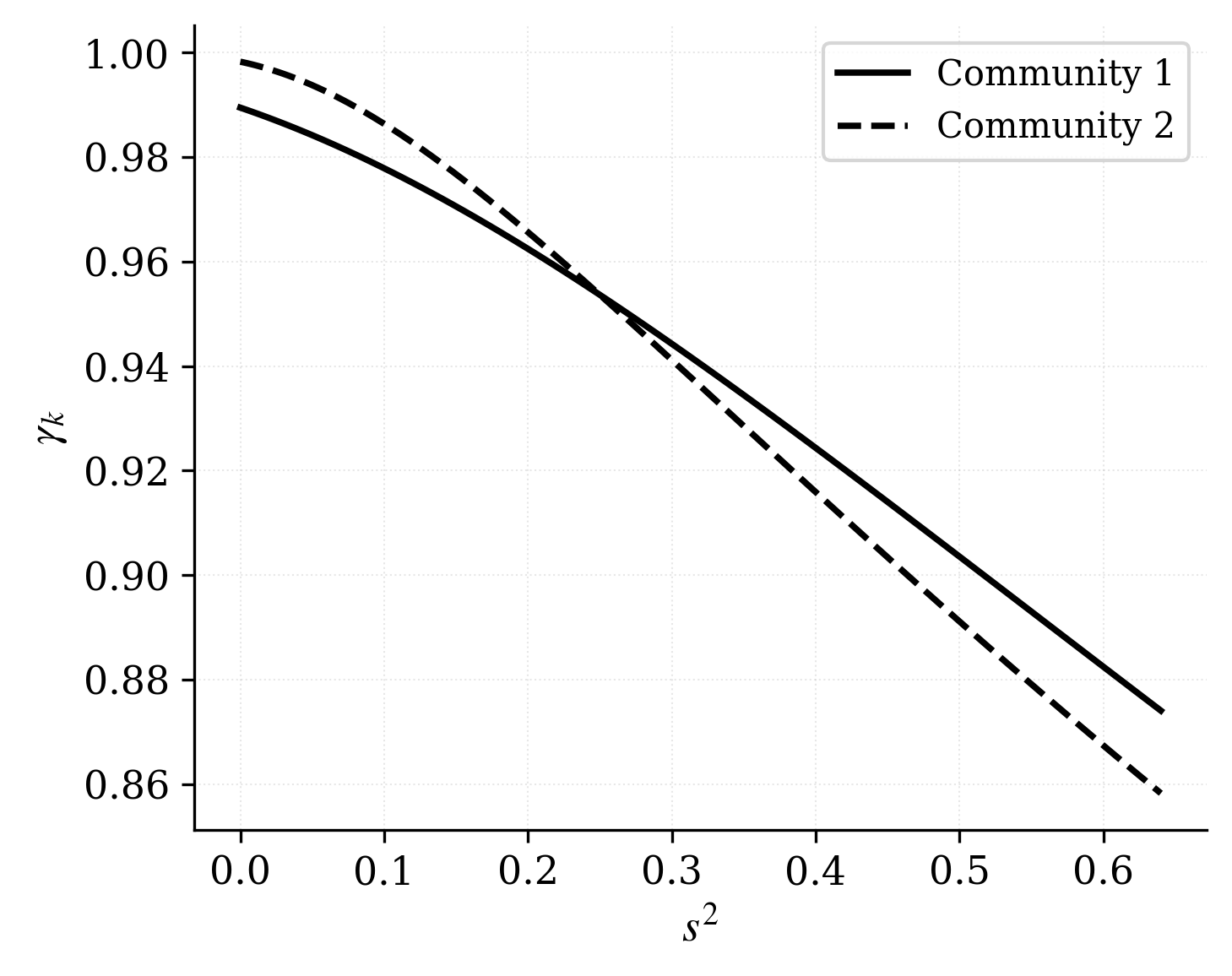}
        \caption{Persistence within each community ($\gamma$) \\ ~}
        \label{fig:gamma_blocks}
    \end{subfigure}
    \hfill
    \begin{subfigure}[b]{0.32\textwidth}
        \centering
        \includegraphics[width=\textwidth]{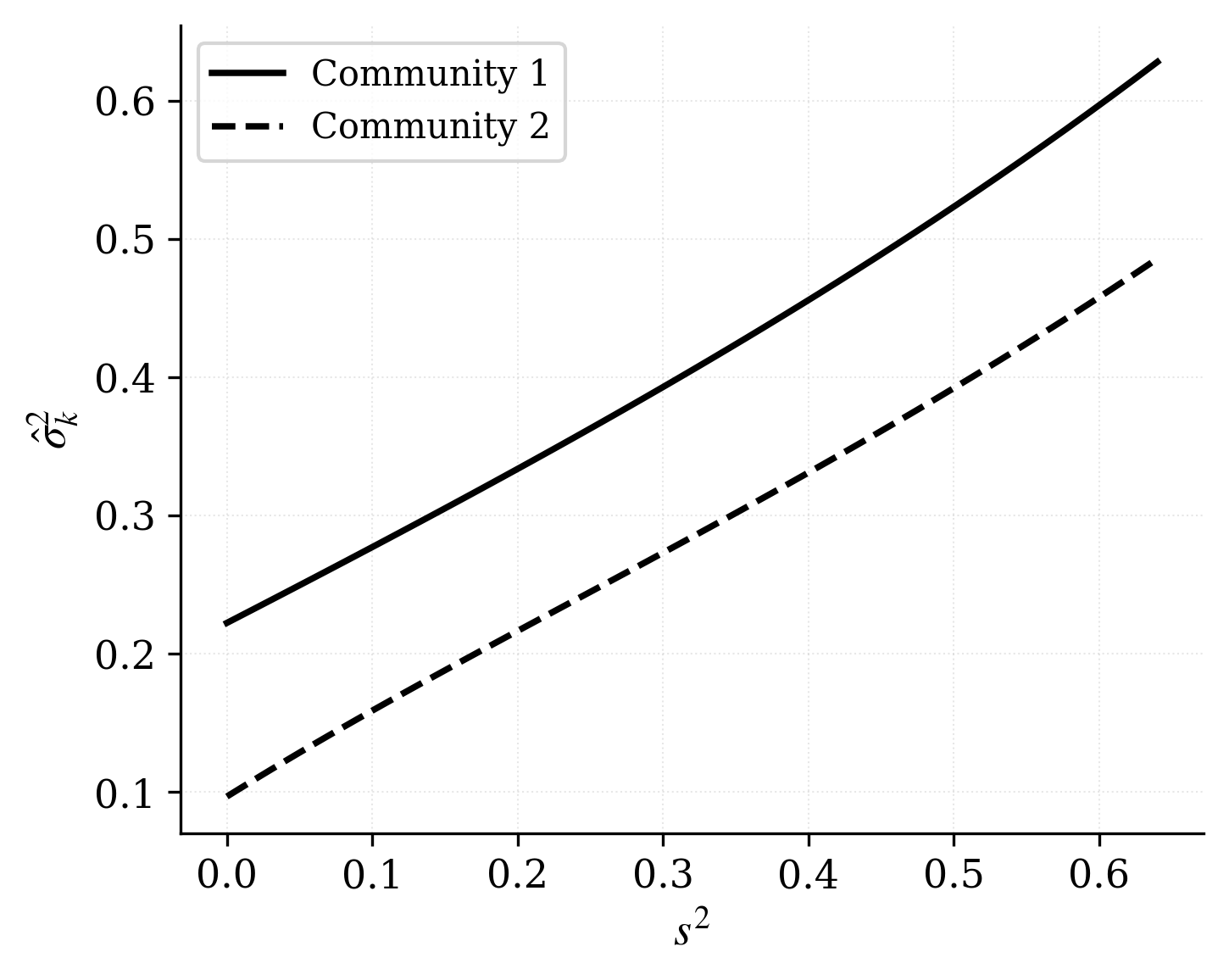}
        \caption{Variance of species abundance within each community ($\sigma^2$)}
        \label{fig:variance_blocks}
    \end{subfigure}
    \hfill
    \begin{subfigure}[b]{0.32\textwidth}
        \centering
        \includegraphics[width=\textwidth]{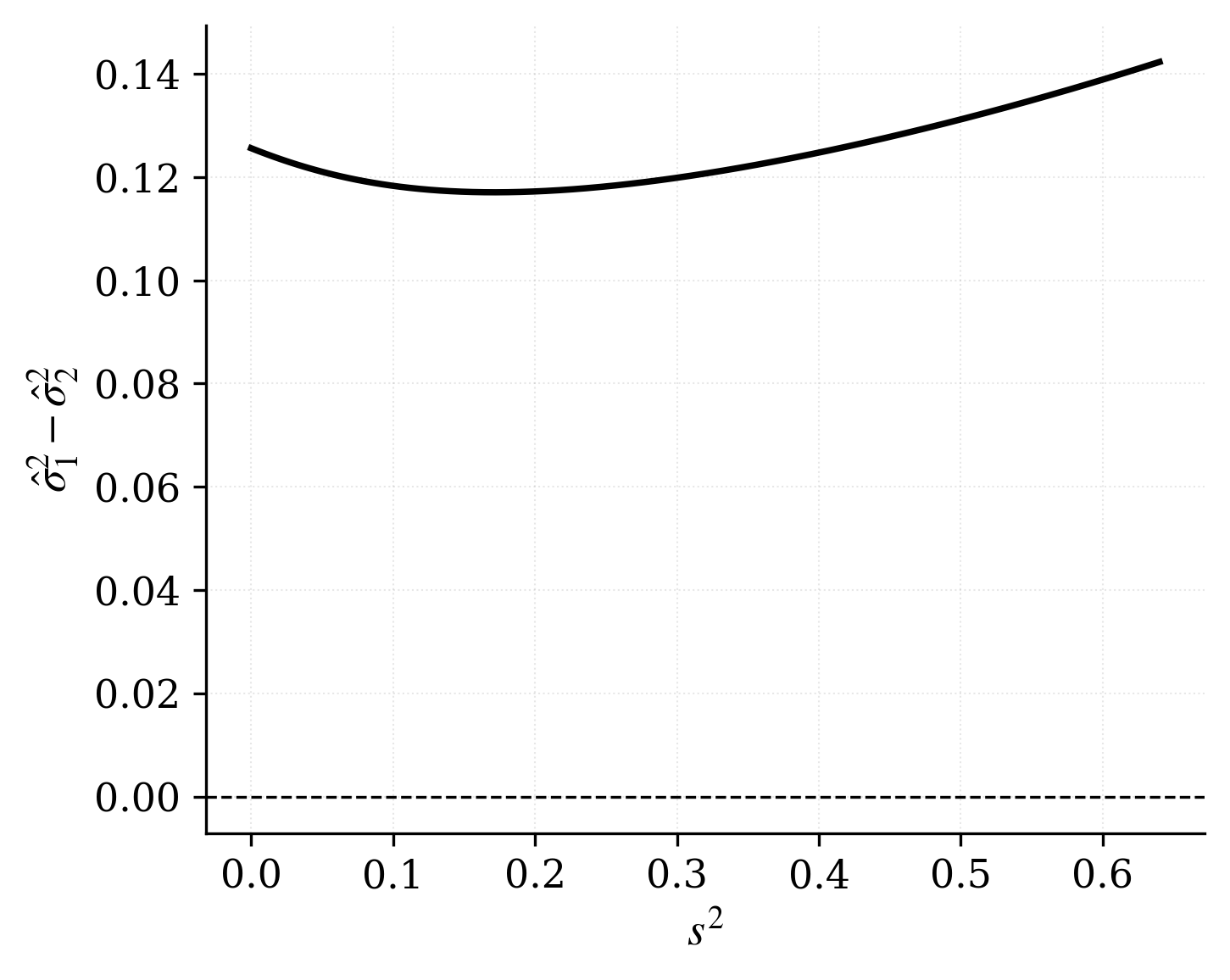}
        \caption{Difference in variance of species abundance between communities}
        \label{fig:variance_difference}
    \end{subfigure}
    \caption{
    Effects of increasing inter-community interaction variance ($s^2 = s^2_{12} = s^2_{21}$) on species persistence and equilibrium abundance variance, under opposing intra-community correlation structures: $\rho_{11} = 0.8$ (mutualistic/competitive) in community 1 and $\rho_{22} = -0.8$ (predator–prey) in community 2.  In panels (a) and (b), the solid line corresponds to community 1 and the dashed line to community 2.
    }
    \label{fig:application2}
\end{figure}

We found that persistence in community 2, decreased more rapidly than in community 1 as inter-community interaction variance increased (Fig.~\ref{fig:gamma_blocks}). When the system reached a point where all interaction variances were equal ($s^2 = 0.25$), both communities have comparable proportion of persisting species (which is consistent with application 1). Beyond this threshold, as the interaction structure became more bipartite, community 1 maintained higher persistence than community 2, despite showing greater variance in equilibrium abundances. Interestingly, while the variance of abundances increased in both communities with stronger inter-community interaction variance (Fig.~\ref{fig:variance_blocks}), the difference in variance between them first decreased (up to approximately $s^2_{12} = 0.25$), then increased again, suggesting a non-monotonic response in variance asymmetry (Fig.~\ref{fig:variance_difference}).

These results highlight how the internal structure of ecological communities can shape their resilience to external interactions. Communities dominated by mutualistic and competitive interactions (positive correlations) appear more resistant to increasing connectivity with other communities, maintaining higher persistence as coupling intensifies. In contrast, communities structured by predator–prey interactions (negative correlations) are more sensitive and prone to collapse. However, this greater persistence comes with more dispersed abundance distributions, which remain broader even at high levels of inter-community interaction. 

\subsection{Feedback between communities under uncorrelated interactions}
In this final application, we investigated whether a feedback loop could emerge in
which increasing the impact of community~$1$ on community~$2$ would ultimately
impact community~$1$ itself. To this end, we set the correlation matrix to zero
($R = 0$) and manipulated the variance of community interactions ($S$). We varied
both directions of interaction variance: $s^2_{12}$ (community~$1$ effect on
community~$2$) and $s^2_{21}$ (the reverse).

\begin{figure}[ht]
    \centering
    \includegraphics[width=0.6\textwidth]{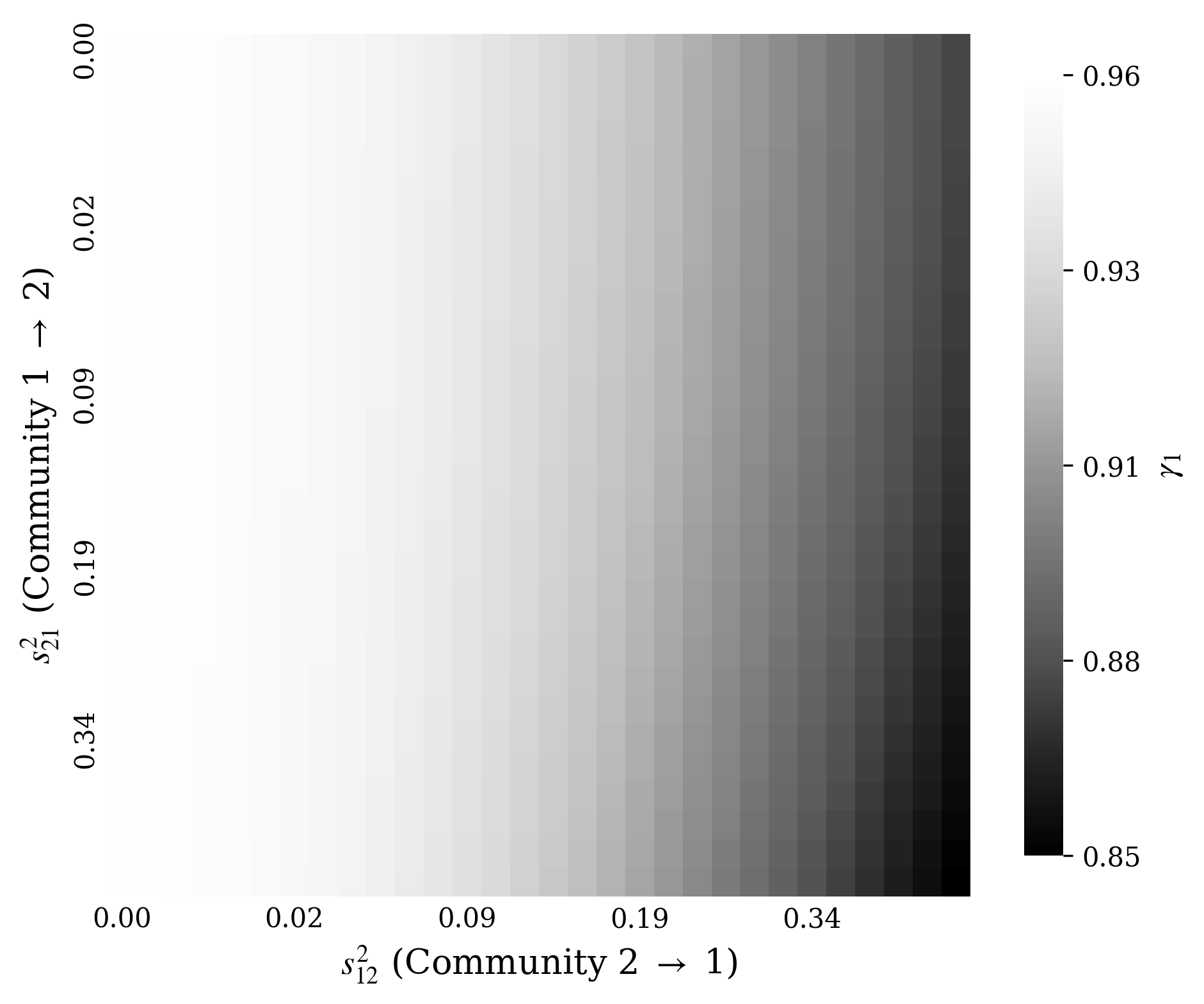}
    \caption{
    Heatmap of the proportion of persisting species in community 1 ($\gamma_1$) as a function of the inter-community interaction variance $s^2_{12}$ and $s^2_{21}$, with zero correlation structure ($R = 0$). The intra-community interaction variance are fixed at $s^2_{11} = s^2_{22} = 0.5$.
    }
    \label{fig:fb_curve}
\end{figure}

As shown in Figure~\ref{fig:fb_curve}, $\gamma_1$ depends jointly and non-separably
on $s^2_{12}$ and $s^2_{21}$: at fixed $s^2_{12}$, $\gamma_1$ stays essentially flat
for small $s^2_{21}$ and drops sharply once $s^2_{21}$ exceeds a threshold that
itself increases with $s^2_{12}$. The persistence of community~$1$ is thus governed
by the reciprocal coupling $s^2_{21}$ in a strongly nonlinear way, even though its
internal structure is held fixed.

We interpret this comparative-statics pattern as an effective feedback between
communities, consistent with a scenario in which a loss of persistence in
community~$2$ amplifies the effect of reciprocal interactions back onto
community~$1$. We stress that this is an interpretation of the stationary
dependence, not a dynamical statement: our analysis characterizes the equilibrium
abundance measure, not a transient trajectory or a temporal cascade. This
emphasizes the importance of maintaining highly persistent communities to prevent
the indirect spread of collapse throughout the ecosystem
\cite{loreau_biodiversity_2001, calcagno_diversity_2017}.

\section{Discussion}

In this paper, we study the LV model, in which species interactions are structured by random interactions with variance and correlation profiles. We examine how this structure of species interactions, which arises from community-level organization, affects significant ecological properties such as persistence and the variance of species abundances at equilibrium.

To study the properties of species abundance at equilibrium, we employed a method based on AMP techniques. Theorem~\ref{thm:main_block} provides an asymptotic characterization of species abundance at equilibrium in the case of a block-structured interaction matrix. Specifically, it shows that the empirical distribution of equilibrium abundances converges, as the number of species $n\to \infty$ to an explicit mixture of truncated Gaussian distributions, whose parameters are determined by the self-consistent fixed-point equations \eqref{FPK}. These equations encode the interplay between the variance profile $S$, the correlation structure $R$, and the block proportions $\alpha$ and jointly determine both the fraction of persisting species within each block and the shape of their abundance distribution. We solved these equations to examine the impact of block-structured correlation and variance profiles on equilibrium in three applications involving two interacting communities. First, we examined the local correlation structure and its impact on global system properties. Then, we explored the impact of intra-community correlations on species persistence when inter-community interactions increase. Finally, we examined feedback between communities in the case of uncorrelated interactions.
\newline

This work contributes to the growing body of results obtained using AMP techniques to analyze equilibrium in large LV systems. Building directly on recent methodological advances \cite{akjouj_equilibria_2024,hachem_approximate_2024,gueddari_elliptic_2025,gueddari_approximate_2026}, this work establishes AMP as a powerful technique for characterizing equilibrium properties beyond classical mean-field approaches.

Historically, LV systems have been investigated from complementary perspectives. In statistical physics, for example, random interaction models have been used to derive generic properties of ecological equilibrium and persistence \cite{bunin_ecological_2017,barbier_generic_2018}, while more recent mathematical works have focused on rigorous characterizations of equilibrium \cite{akjouj_complex_2024}. Within this literature, the role of correlation between interaction coefficients has been explored, notably in relation to feasibility and stability \cite{clenet_equilibrium_2022,poley_generalized_2023}. However, these studies have thus far been limited to homogeneous correlation structures. A key contribution of this work is overcoming this limitation by introducing correlation profiles. This extension reveals new qualitative effects that cannot be captured by homogeneous correlation alone.

Application 1 reveals that introducing correlation within a single community while maintaining homogeneous interaction variance can significantly impact the persistence of species across the entire ecosystem (Fig.~\ref{fig:2}) This result is consistent with earlier findings showing that correlation can influence ecological stability and persistence \cite{bunin_ecological_2017}, but it contrasts sharply with previous results on feasibility thresholds, where correlation was found to have no effect on the threshold \cite{clenet_equilibrium_2022}. 

Finally, this work naturally extends recent efforts on block-structured LV models \cite{barbier_generic_2018,clenet_impact_2024}. Application 2 shows that communities dominated by positive interaction correlations, which are typical of mutualistic or competitive systems, are more robust to increasing inter-community interactions. In contrast, communities structured by negative correlations, such as those in predator-prey interactions, become markedly more sensitive and prone to collapse as inter-community interactions intensify (Fig.~\ref{fig:application2}). Application 3 shows that an effective, feedback-like dependence of one community's persistence on the reciprocal coupling can arise from asymmetries in interaction variance alone, in the complete absence of correlation (Fig.~\ref{fig:fb_curve}). This result emphasizes that structural heterogeneity alone can generate effective feedback, further enriching the spectrum of mechanisms that shape persistence and stability in complex ecological systems \cite{barbier_generic_2018}.
\newline

Although the present framework can capture new structural effects in large LV systems, it relies on several simplifying assumptions that naturally limit its scope.

One limitation stems from the class of correlation structures considered. In this work, correlations are imposed through structured profiles acting at the block level; however, they remain restricted to relatively simple patterns. More general correlation architectures could be explored, such as correlations acting along rows or columns or directly within inter-community interaction matrices \cite{castedo_generalized_2025}. Similarly, correlations that are not diagonally opposed, such as asymmetric or direction-dependent patterns, remain outside the scope of the current framework \cite{baron_eigenvalues_2022}. Extending AMP to include these heterogeneous and anisotropic correlation structures would greatly increase its ecological relevance.

Our focus on zero-mean interaction matrices implicitly neglected the role of the average interaction strength $\mu$. Yet, $\mu$ is known to be a key control parameter in LV systems, governing the balance between predominantly mutualistic or predominantly competitive communities \cite{clenet_equilibrium_2022,clenet_equilibrium_2023}. 

Another limitation is the level at which correlation is characterized. While our analysis defines correlation at the level of the interaction matrix, it does not explicitly address how the dynamic itself reshapes correlations. For example, the interaction matrix restricted to surviving species may exhibit correlation patterns that differ markedly from those of the original system. Studying these emergent correlations would align with recent work on ecological fingerprints and post-assembly interaction structures \cite{barbier_fingerprints_2021,poley_interaction_2025}, and could provide deeper insight into the feedback between assembly, extinction, and interaction structure.
\newline

Our results offer new perspectives in theoretical ecology. One promising direction is the extension to multilayer or multiplex formulations, in which each block or layer represents a distinct ecological dimension. Examples of these dimensions include interaction type (e.g., trophic versus mutualistic), spatial landscape, and temporal snapshots \cite{pilosof_multilayer_2017,wang_stability_2023}. By modulating the variance and correlation of interactions across layers, one can investigate the propagation of stabilizing or destabilizing effects through structured ecological networks and the influence of asymmetries in coupling among layers on collective dynamics. From this broader perspective, the classical concept of keystone species \cite{paine_food_1966,paine_pisaster-tegula_1969} could be extended to keystone communities \cite{mouquet_extending_2013}.

Our analysis of emergent feedback loops between two communities highlights the importance of generalizing to systems with three or more interacting communities. In such cases, indirect feedback could arise. For instance, perturbations in community 1 could influence community 2, thereby affecting community 3. Ultimately, this feeds back onto the initial community. These multi-loop feedback structures, or intransitive interaction motifs, resemble the well-known ``rock–paper–scissors" dynamics observed in ecological systems, where competitive dominance follows a cyclic rather than hierarchical pattern. These configurations can promote long-term coexistence by preventing any single community from achieving total dominance \cite{allesina_competitive_2011,grilli_keystone_2023}. However, these dynamics have yet to be explored within a community-type framework.

Finally, although this study is theoretical by intention, it paves the way for future research at the intersection of theory, inference, and data. A key challenge will be connecting AMP-based predictions to empirically inferred ecological networks. Recent progress in microbial ecology, where interaction structures are reconstructed from abundance data and time series, provides a natural context in which to explore these ideas \cite{hu_emergent_2022}.

\section*{Fundings}
\noindent M.C. acknowledge financial support from the Institut Quantique at Université de Sherbrooke and the FRQNT .

\section*{Data availability} 
\noindent There is no data associated with this manuscript.

\section*{Conflict of interest disclosure}
\noindent The authors declare that they have no financial conflict of interest with the content of this article.

\section*{Numerical simulations}
\noindent Simulations were performed in Python. All the figures and the code are available on Github \cite{code}.

\newpage

\vfill\pagebreak
\small
\printbibliography

\newpage

\appendix

\section{Proof outline of Theorem~\ref{thm:main_block}}
\label{app:proof}

\subsection{Asymptotics of the empirical measure}

For the sake of generality, we outline the main elements of the proof within the framework of Conjecture~\ref{conj:general}. Our strategy follows the approach developed in \cite[Section~5]{hachem_approximate_2024}, and we therefore only sketch the principal arguments.

The main novelty of our setting lies in the presence of a correlation profile in addition to the variance profile. To incorporate this additional structure, we rely on the AMP results established in \cite{gueddari_approximate_2026}, which extend the theory to matrices with general variance and correlation profiles.

We first observe that the equilibrium vector $u^\star$ solves the linear complementarity problem
$$u^\star = \operatorname{LCP}(I_n - \Sigma, -r).$$
We design an AMP algorithm to approximate the solution of this LCP. Let $\tilde S \in \mathbb{R}^{n \times n}$ be a deterministic matrix (to be specified later). Initialize $x^0 = \mathbf{1} \in \mathbb{R}^n$ and consider the activation function
$$h(x,\eta) = (x+\eta)_+, \qquad x,\eta \in \mathbb{R}.$$
Let $(x^t)$ be generated by $(x^t) = \operatorname{AMP}(X,h,x^0,\tilde S)$, that is,
\begin{equation}
\label{eq:AMP}
x^{t+1} = W (x^t+\eta)_+ - \Big(\mathcal{\tilde S}\, \mathbb{P}(Z^t+\eta \ge 0)\Big) \odot (x^{t-1}+\eta)_+,
\end{equation}
where
$$W = \tilde S^{\odot 1/2} \odot X, \qquad \mathcal{\tilde S} = (\tilde S \odot \tilde S^\top)^{\odot 1/2} \odot T,$$
and $(Z^1,\dots,Z^t)$ are $n$-dimensional centered Gaussian vectors whose covariances satisfy the Density Evolution equations (see \cite[Section~1.3]{gueddari_approximate_2026}).

Let $a^t = (a_i^t) \in \mathbb{R}^n$ denote the vector of variances, $a_i^t = \mathbb{E}(Z_i^t)^2$, so that $(a^t)$ satisfies the recursion
$$\begin{cases} a^1 = \tilde S (1+\eta)^2, \\ a^{t+1} = \tilde S \, \mathbb{E}\!\left(\sqrt{a^t}\xi + \eta\right)_+^2. \end{cases}$$
Define also $\zeta^t = \mathcal{\tilde S}\, \mathbb{P}\!\left(\sqrt{a^t}\xi + \eta \ge 0\right)$. Assume that $(a^t)$ and $(\zeta^t)$ converge to $a \in \mathbb{R}^n$ and $\zeta \in \mathbb{R}^n$. Then $(a,\zeta)$ solves
\begin{equation}
\label{eq:FPE1}
a = \tilde S\, \mathbb{E}\!\left(\sqrt{a}\xi+\eta\right)_+^2, \qquad \zeta = \mathcal{\tilde S}\, \mathbb{P}\!\left(\sqrt{a}\xi+\eta \ge 0\right).
\end{equation}

For large $t$, we approximate $x^{t+1} \approx x^t \approx x^{t-1} = x$ and $\zeta^t \approx \zeta$. Then \eqref{eq:AMP} becomes
$$x = W(x+\eta)_+ - \zeta \odot (x+\eta)_+.$$
After straightforward manipulations, this leads to the observation that $u = (1+\zeta)^{1/2}(x+\eta)_+$ solves the LCP
$$\operatorname{LCP}\!\left( I_n - (1+\zeta)^{-1/2} W (1+\zeta)^{-1/2}, -(1+\zeta)^{-1/2}\eta \right).$$
We therefore choose
$$W = (1+\zeta)^{1/2} \Sigma (1+\zeta)^{1/2}, \qquad \eta = (1+\zeta)^{1/2} r,$$
which implies $\tilde S = (1+\zeta) V (1+\zeta)$, and similarly, with $\mathcal V = (V \odot V^\top)^{\odot 1/2} \odot T$, that $\mathcal{\tilde S} = (1+\zeta) \mathcal V (1+\zeta)$. Substituting into \eqref{eq:FPE1}, we obtain the $2n$-dimensional system
\begin{equation}
\label{eq:FPE22}
\begin{aligned}
a &= (1+\zeta)V(1+\zeta)\, \mathbb{E}\!\left( \sqrt{a}\xi + (1+\zeta)^{1/2}r \right)_+^2, \\
\zeta &= (1+\zeta)\mathcal V(1+\zeta)\, \mathbb{P}\!\left( \sqrt{a}\xi + (1+\zeta)^{1/2}r \ge 0 \right).
\end{aligned}
\end{equation}
Introducing the change of variables $p = \frac{a}{1+\zeta}$, we obtain the equivalent system
$$\begin{cases} p = V\,\mathrm{diag}(1+\zeta)^2 \mathbb{E}\!\left(\sqrt{p}\xi + r\right)_+^2, \\[0.2cm] \zeta = \mathrm{diag}(1+\zeta) \mathcal V \mathrm{diag}(1+\zeta) \mathbb{P}\!\left(\sqrt{p}\xi+r \ge 0\right), \end{cases}$$
with $(p,\zeta)\in \mathbb{R}_+^n \times [-1,1]^n$.

The AMP result implies that, in the high-dimensional regime and as $t \to \infty$, the iterates $(x^t)$ behave like a Gaussian vector $Z^\infty \in \mathbb{R}^n$ whose $i$-th coordinate has variance $a_i$. Recalling that the vector approximating the equilibrium is
$$u^t = (1+\zeta)^{1/2} \big( x^t + (1+\zeta)^{1/2} r \big)_+,$$
we obtain the claimed limiting empirical distribution.

\subsection{Existence of a solution to the fixed point equations}

The parameters describing the asymptotic empirical measure of the equilibrium solve a system of fixed-point equations, and one must show that this system admits a solution. We establish this in the block-structured case, namely we show that the system \eqref{FPK} admits a solution. The argument relies on Brouwer's fixed-point theorem. For simplicity we take $r=1$.

Let $A$ and $B$ be the $K \times K$ matrices defined by
$$A_{ij} = \alpha_j \rho_{ij} s_{ij} s_{ji}, \qquad B_{ij} = \alpha_j s_{ij}^2.$$
The system can then be rewritten as
$$\left\{ \begin{aligned} \delta_i &= [f(\delta, \sigma)]_i := 1 - \sum_{j=1}^K A_{ij} \frac{\mathbb{P}\!\left( \sigma_j \xi + 1 \ge 0 \right)}{\delta_j}, \qquad i=1,\dots,K, \\[0.2cm] \sigma_i^2 &= [g(\delta, \sigma)]_i := \sum_{j=1}^K B_{ij} \frac{\mathbb{E}\!\left( \sigma_j \xi + 1 \right)_+^2}{\delta_j^2}, \qquad i=1,\dots,K. \end{aligned} \right.$$

Since $|\rho_{ij}|\le 1$ and $2\, s_{ij} s_{ji} \le s_{ij}^2 + s_{ji}^2$, we have
$$\|A\|_\infty \le \tfrac12\big(\|S\|_\infty + \|S^\top\|_\infty\big), \qquad \|B\|_\infty \le \|S\|_\infty.$$
In particular, when $S$ is symmetric, the assumption $\|S\|_\infty \le \tfrac18$ gives $\|A\|_\infty \le \tfrac18$ and $\|B\|_\infty \le \tfrac18$ (for a non-symmetric $S$ it suffices to assume in addition $\|S^\top\|_\infty \le \tfrac18$). We assume this from now on.

To apply Brouwer's theorem it suffices to exhibit a compact convex set invariant under the continuous map
$$(\delta,\sigma) \mapsto \big( f(\delta,\sigma),\ \sqrt{g(\delta,\sigma)} \big).$$
We claim that the set $[1/2,\,3/2]^K \times [0,\,1]^K$ is invariant. Let $(\delta,\sigma)$ belong to this set, so that $1/\delta_j \le 2$ and $\mathbb{P}(\sigma_j \xi + 1 \ge 0) \le 1$.

For the first equation,
$$\big| [f(\delta,\sigma)]_i - 1 \big| \le \sum_{j=1}^K |A_{ij}| \frac{1}{\delta_j} \le 2 \sum_{j=1}^K |A_{ij}| \le 2\|A\|_\infty \le \tfrac14,$$
hence $[f(\delta,\sigma)]_i \in [3/4,\, 5/4] \subset [1/2,\, 3/2]$.

For the second equation, using $\mathbb{E}(\sigma_j \xi + 1)_+^2 \le 1 + \sigma_j^2 \le 2$ and $1/\delta_j^2 \le 4$,
$$[g(\delta,\sigma)]_i \le 4 \sum_{j=1}^K B_{ij}\,(1+\sigma_j^2) \le 8 \sum_{j=1}^K B_{ij} \le 8\|B\|_\infty \le 1,$$
hence $\sqrt{[g(\delta,\sigma)]_i} \le 1$.

Therefore the map sends $[1/2,3/2]^K \times [0,1]^K$ into itself. This set is compact and convex, and the map is continuous, so Brouwer's fixed-point theorem yields at least one solution of the system \eqref{FPK}.

\section{Numerical validation of Conjecture~\ref{conj:general}}
\label{app:conj_num}

Theorem~\ref{thm:main_block} and Figure~\ref{fig:lcp_mixture} concern the block-structured model, where the limiting fixed-point system collapses to the $2K$ equations~\eqref{FPK}. We now provide numerical evidence that the characterization of Conjecture~\ref{conj:general} also holds for a non-block profile, meaning when the variance and correlation vary continuously across species.

We take $n = 2000$, $r = \mathbf{1}$, and place the species on a grid $x_i = (i-\tfrac12)/n \in (0,1)$. The variance profile is chosen \emph{row-dependent}, hence asymmetric ($v_{ij}\neq v_{ji}$), and the correlation
profile is sign-varying along the spectrum:
$$
v_{ij} = \frac{s_0^2\,(1 + c\,x_i)}{n},
\qquad
\tau_{ij} = \rho_0 \cos\!\big(\pi(x_i + x_j)\big),
$$
with $s_0 = 0.55$, $c = 0.6$, $\rho_0 = 0.5$. Neither profile is piecewise
constant, so this configuration lies strictly outside the scope of
Theorem~\ref{thm:main_block}; the correlation profile in particular ranges
continuously from predator--prey ($\tau_{ij} < 0$) to competitive/mutualistic
($\tau_{ij} > 0$) pairs.

To make this profile concrete, evaluating it on the coarse grid $n=3$, where $x=(\tfrac16,\tfrac12,\tfrac56)$, gives a row-constant variance matrix and a sign-varying correlation matrix
\[
V=\begin{pmatrix}0.111&0.111&0.111\\0.131&0.131&0.131\\0.151&0.151&0.151\end{pmatrix},
\qquad
T=\begin{pmatrix}\phantom{-}0.25&-0.25&-0.50\\-0.25&-0.50&-0.25\\-0.50&-0.25&\phantom{-}0.25\end{pmatrix}.
\]
The row-dependence of $V$ is visible as the increasing interaction strength from the first to the last species, while $T$ interpolates from competitive/mutualistic pairs on the diagonal corners ($\tau_{11}=\tau_{33}=0.25$) to predator--prey coupling at the anti-diagonal ($\tau_{13}=-0.5$).

\begin{figure}[ht]
    \centering
    \includegraphics[width=0.62\textwidth]{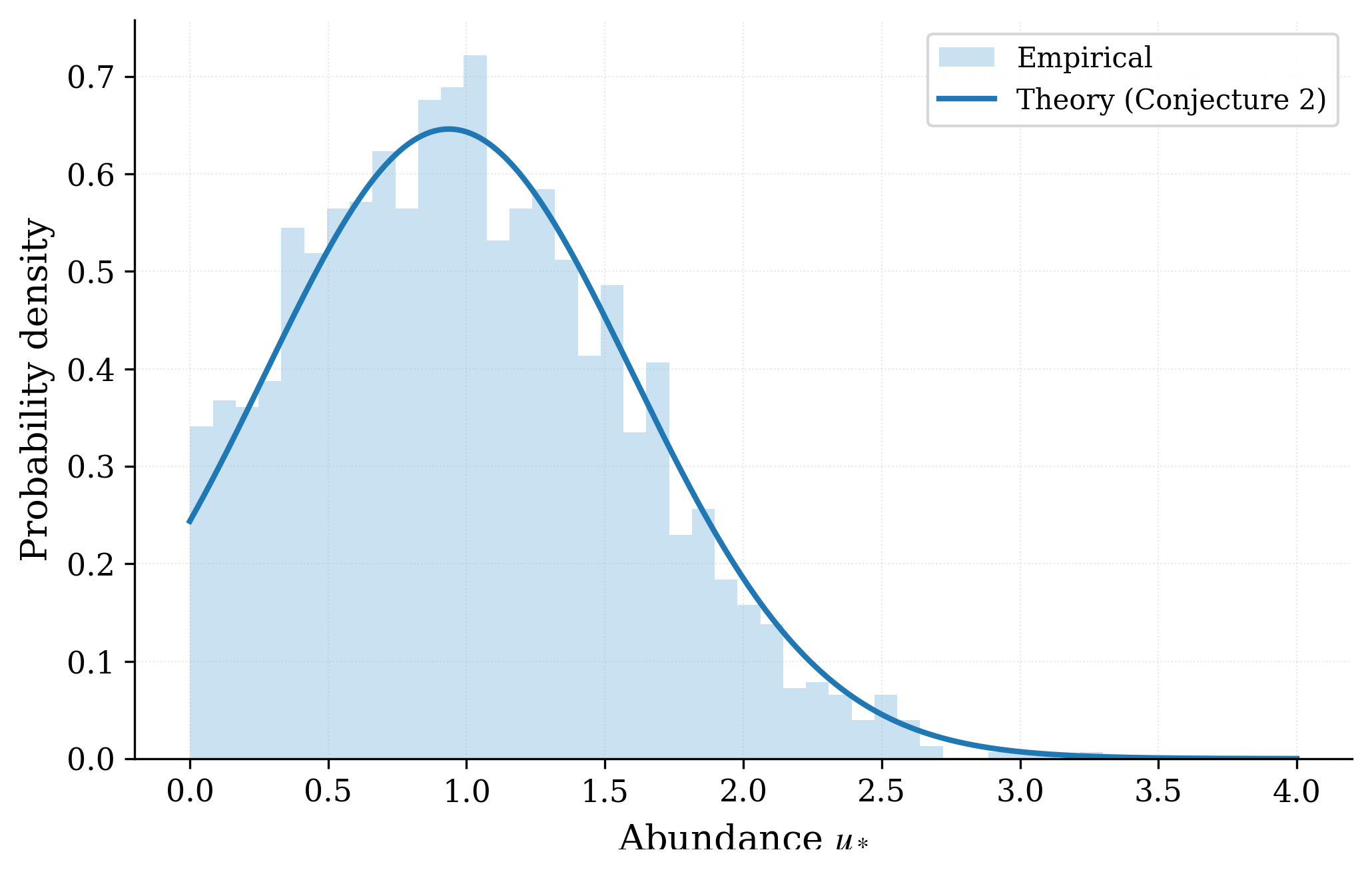}
    \caption{
    Numerical validation of Conjecture~\ref{conj:general} for a continuous,
    non-block profile: row-dependent (asymmetric) variance
    $v_{ij} = s_0^2(1+c\,x_i)/n$ and sign-varying correlation
    $\tau_{ij} = \rho_0\cos(\pi(x_i+x_j))$, with $x_i=(i-\tfrac12)/n$,
    $s_0 = 0.55$, $c = 0.6$, $\rho_0 = 0.5$, $n = 2{,}000$, $r = \mathbf 1$. The histogram shows the empirical abundance distribution of the surviving species, and the solid line the density predicted by Conjecture~\ref{conj:general}. 
    }
    \label{fig:conj_validation}
\end{figure}

We solve the $2n$-dimensional system~\eqref{eq:FPE2} by iterating the associated
density-evolution recursion to convergence, which yields $(p,\zeta)$ and hence the
predicted measure $\mu_n = \mathcal{L}\big((Y_\theta)_+\big)$ with
$Y_i = (1+\zeta_i)(\sqrt{p_i}\,\xi + r_i)$. On the empirical side, we sample the
interaction matrix $\Sigma$ from the above profile and solve the linear
complementarity problem $u^\star = \operatorname{LCP}(I_n - \Sigma, -r)$ directly.
Figure~\ref{fig:conj_validation} shows that the density of the surviving species
predicted by Conjecture~\ref{conj:general} matches the empirical histogram.

\end{document}